\documentclass[11pt,tightenlines,aps,prd,superscriptaddress,nofootinbib,nobibnotes,longbibliography,notitlepage]{revtex4-1}
\pdfoutput=1

\usepackage{graphicx}
\usepackage{amsfonts}
\usepackage{amssymb}
\usepackage{amsbsy}
\usepackage{amsmath}
\usepackage{mathtools}
\usepackage{latexsym}
\usepackage{bm}
\usepackage{color}
\usepackage{hyperref}
\usepackage[margin=2cm]{geometry}
\usepackage{comment}
\usepackage[makeroom]{cancel}
\usepackage[dvipsnames]{xcolor}
\usepackage{empheq}

\usepackage{stackengine}
\stackMath
\newcommand\tenq[2][1]{%
 \def\useanchorwidth{T}%
  \ifnum#1>1%
    \stackon[0pt]{\tenq[\numexpr#1-1\relax]{#2}}{\scriptscriptstyle\sim}%
  \else%
    \stackon[1pt]{#2}{\scriptscriptstyle\sim}%
  \fi%
}

\newcommand{\de}{\mbox{d}}
\newcommand{\lf}{\left}
\newcommand{\rg}{\right}
\newcommand{\be}{\begin{equation}}
\newcommand{\ee}{\end{equation}}

\newcommand{\bea}{\begin{eqnarray}}
\newcommand{\eea}{\end{eqnarray}}

\numberwithin{equation}{section}

\renewcommand{\theequation}{\arabic{section}.\arabic{equation}}

\begin{document}

\title{A low-redshift preference for an interacting dark energy model}

\author{Yuejia Zhai}
\email{y.zhai@sheffield.ac.uk}
\affiliation{School of Mathematical and Physical Sciences, University of Sheffield, Hounsfield Road, Sheffield S37RH, United Kingdom}

\author{Marco de Cesare}
\email{marco.decesare@na.infn.it}
\affiliation{Scuola Superiore Meridionale,
Largo San Marcellino 10, 80138 Napoli, Italy}
\affiliation{INFN, Sezione di Napoli, Monte S. Angelo, Via Cintia, 80126 Napoli, Italy}

\author{Carsten van de Bruck}
\email{c.vandebruck@sheffield.ac.uk}
\affiliation{School of Mathematical and Physical Sciences, University of Sheffield, Hounsfield Road, Sheffield S37RH, United Kingdom}

\author{Eleonora Di Valentino}
\email{e.divalentino@sheffield.ac.uk}
\affiliation{School of Mathematical and Physical Sciences, University of Sheffield, Hounsfield Road, Sheffield S37RH, United Kingdom}

\author{Edward Wilson-Ewing}
\email{edward.wilson-ewing@unb.ca}
\affiliation{Department of Mathematics and Statistics, University of New Brunswick, Fredericton, NB, E3B 5A3, Canada}
\affiliation{Department of Physics, McGill University, Montr\'eal, QC, H3A 2T8, Canada}

\begin{abstract}
We explore an interacting dark sector model in trace-free Einstein gravity where dark energy has a constant equation of state, $w=-1$, and the energy-momentum transfer potential is proportional to the cold dark matter density. Compared to the standard $\Lambda$CDM model, this scenario introduces a single additional dimensionless parameter, $\epsilon$, which determines the amplitude of the transfer potential. 
Using a combination of \textit{Planck} 2018 Cosmic Microwave Background (CMB), DESI 2024 Baryon Acoustic Oscillation (BAO), and Pantheon+ Type Ia supernovae (SNIa) data, we derive stringent constraints on the interaction, finding $\epsilon$ to be of the order of $\sim \mathcal{O}(10^{-4})$. While CMB and SNIa data alone do not favor the presence of such an interaction, the inclusion of DESI data introduces a mild $1\sigma$ preference for an energy-momentum transfer from dark matter to dark energy. This preference is primarily driven by low-redshift DESI BAO measurements, which favor a slightly lower total matter density $\Omega_m$ compared to CMB constraints.
Although the interaction remains weak and does not significantly alleviate the $H_0$ and $S_8$ tensions, our results highlight the potential role of dark sector interactions in late-time cosmology. 
\end{abstract}

\maketitle

\nopagebreak

\section{Introduction}

The Planck mission has marked the beginning of an exciting decade for cosmology, with its high-precision measurements of the cosmic microwave background~\cite{Planck:2013oqw, Planck:2013pxb, Planck:2018vyg}. The current understanding of the evolving Universe is built on a straightforward assumption of its homogeneity and isotropy, plus small inhomogeneous perturbations. Starting from this cornerstone, the widely accepted standard cosmological paradigm is the $\Lambda$CDM model based on general relativity, which assumes that dark matter is cold (CDM), single scalar-field inflation, and the existence of a cosmological constant ($\Lambda$) as the theoretical basis for the Universe's accelerated expansion. The $\Lambda$CDM model has been remarkably successful in explaining observations ranging from temperature and polarization anisotropies in the cosmic microwave background (CMB)~\cite{Planck:2018vyg,ACT:2020gnv,SPT-3G:2021eoc,SPT-3G:2022hvq}, to the large-scale structure of the Universe~\cite{DES:2017qwj,eBOSS:2020yzd}; however, concerns about the $\Lambda$CDM model have been raised from both theoretical and observational perspectives since the model was first developed~\cite{Perivolaropoulos:2021jda,Abdalla:2022yfr,DiValentino:2022fjm,Peebles:2024txt}. On one hand, the $\Lambda$CDM model suffers from the fine-tuning and coincidence problems~\cite{Weinberg:1988cp}. The fine-tuning problem is that, assuming the $\Lambda$CDM model in data analysis, the derived cosmological constant is 54 orders of magnitude smaller than the vacuum energy predicted by quantum theory~\cite{Martin:2012bt}, while the coincidence problem is that today's energy density of total mass in the Universe, $\Omega_m$, coincides in order of magnitude with that of the vacuum energy, $\Omega_\Lambda$. On the other hand, the cosmological parameters measured from recent CMB observations disagree with those measured by local experiments at a $2$--$5\sigma$ level. Among these tensions, the discrepancies in the Hubble constant $H_0$~\cite{Verde:2019ivm,DiValentino:2020zio,DiValentino:2021izs,Schoneberg:2021qvd,Shah:2021onj,Kamionkowski:2022pkx,Giare:2023xoc,Hu:2023jqc,Verde:2023lmm,DiValentino:2024yew,Perivolaropoulos:2024yxv} and the matter clustering parameter $S_8 \equiv \sigma_8 (\Omega_m/0.3)^{1/2}$~\cite{DiValentino:2020vvd,DiValentino:2018gcu,DES:2021bvc,DES:2021vln,KiDS:2020suj,Asgari:2019fkq,Joudaki:2019pmv,DAmico:2019fhj,Kilo-DegreeSurvey:2023gfr,Troster:2019ean,Heymans:2020gsg,Dalal:2023olq,Chen:2024vvk,ACT:2024okh,DES:2024oud,Harnois-Deraps:2024ucb,Dvornik:2022xap,DES:2021wwk} stand out as the most significant.

Interacting dark energy (IDE) models have been proposed as a possible solution to these tensions~\cite{Kumar:2016zpg, Murgia:2016ccp, Kumar:2017dnp, DiValentino:2017iww,Bahamonde:2017ize,Kumar:2021eev,Pan:2023mie,Benisty:2024lmj,Yang:2020uga,Forconi:2023hsj,Pourtsidou:2016ico,DiValentino:2020vnx,DiValentino:2020leo,Nunes:2021zzi,Yang:2018uae,vonMarttens:2019ixw,Lucca:2020zjb,Zhai:2023yny,Bernui:2023byc,Hoerning:2023hks,Giare:2024ytc,Escamilla:2023shf,vanderWesthuizen:2023hcl,Silva:2024ift,DiValentino:2019ffd,Li:2024qso,Pooya:2024wsq,Halder:2024uao,Castello:2023zjr,Yao:2023jau,Mishra:2023ueo,Nunes:2016dlj,Teixeira:2024qmw}. Although certain phenomenological interactions between dark matter and dark energy have been found to relieve the tensions~\cite{DiValentino:2019ffd}, they suffer from criticisms such as large-scale instability and dependence on non-local variables~\cite{Boehmer:2008av}, like the Hubble parameter, which can be seen as a consequence of the first principle of thermodynamics~\cite{Nunes:2022bhn}.

Alternatively, the trace-free Einstein equations provide a minimal modification to the usual Einstein equations that naturally allows IDE models where dark energy has a constant equation of state $w=-1$ (similar to interacting vacuum models), and the energy-momentum transfer between dark matter and dark energy originates from a potential~\cite{Josset:2016vrq, Perez:2020cwa, deCesare:2021wmk}. Importantly, these IDE models are not affected by the large-scale instability that plagues many IDE models~\cite{Valiviita:2008iv, deCesare:2021wmk}.

In general, the conservation of the matter stress-energy tensor does not follow from the trace-free Einstein equations but rather represents an independent assumption that leads to unimodular gravity, from which general relativity can be recovered with $\Lambda$ playing the role of a free integration constant~\cite{Ellis:2010uc}. Instead, by allowing violations of the conservation of the matter stress-energy tensor, in the case where the energy-momentum loss is integrable, it is possible to recast the dynamics in the form of the usual Einstein field equations; however, now $\Lambda$ is not a constant but rather a spacetime function~\cite{Josset:2016vrq, Perez:2020cwa}. In this case, the trace-free Einstein equations must be complemented by a specific model for the energy-momentum transfer, which determines the evolution of $\Lambda$.

Within this context, it has been argued that energy-momentum non-conservation may arise due to diffusive interactions of low-energy matter degrees of freedom with the discrete quantum-gravity structures underlying spacetime~\cite{Josset:2016vrq}. Such non-conservation could potentially have an impact on cosmological scales, for example as a source of dark energy~\cite{Perez:2017krv, Perez:2018wlo}, in alleviating the Hubble tension~\cite{Perez:2020cwa}, the $S_8$ tension \cite{Albertini:2024yfm}, and on the CMB~\cite{Landau:2022mhm}.

In this paper, we derive constraints from cosmological datasets for a simple IDE model where the transfer potential is a linear function of the energy density of cold dark matter. Compared to $\Lambda$CDM, the model features an additional dimensionless free parameter $\epsilon$, which represents the coupling strength between dark matter and dark energy and leads to an effective non-zero sound speed of dark matter (due to this effect, the model can equally be understood as a generalized dark matter model~\cite{Hu:1998kj, Kopp:2016mhm, Thomas:2016iav, Ilic:2020onu}). Requiring the absence of gradient instabilities ensures that energy flows from dark matter to dark energy, restricting $\epsilon$ to non-negative values.

The paper is organized as follows. In Sec.~\ref{Sec:Model}, we review the formulation of the cosmological model at hand and its embedding within trace-free Einstein gravity. In Sec.~\ref{Sec:Methodology}, we discuss the methodology used for the numerical simulations and data analysis. Our results are presented in Sec.~\ref{Sec:Results}, where we discuss constraints on the parameters of the model. Finally, in Sec.~\ref{Sec:Conclusions}, we conclude and summarize our findings. A supplementary Appendix~\ref{Sec:AppendixA} is also included, where we provide the dynamical equations for the cosmological background and perturbations in the synchronous gauge for a more general model with an arbitrary transfer potential.

\section{Interactions in the dark sector}
\label{Sec:Model}

The trace-free Einstein equations
\be\label{Eq:TraceFreeEFEs}
R_{ab} - \frac{1}{4} R \, g_{ab} = \kappa \left( T_{ab} - \frac{1}{4} T \, g_{ab} \right)
\ee
are closely related to the standard Einstein equations, although with the important difference that, in this formulation, the stress-energy tensor is not necessarily conserved~\cite{Ellis:2010uc}. Here, $\kappa \equiv 8\pi G$ denotes the gravitational coupling.
The energy-momentum transfer
\be
J_a = \kappa \nabla^b T_{ab}
\ee
measures how strongly the conservation of the stress-energy tensor is violated. An interesting case occurs when the energy-momentum transfer is integrable, that is, $J_a = \nabla_a Q$ for some function $Q$, in which case it is possible to rewrite the trace-free Einstein equations as~\cite{Josset:2016vrq,Perez:2020cwa,deCesare:2021wmk}
\be \label{Eq:EffectiveEFE}
G_{ab} = \kappa T_{ab} + Q \, g_{ab}~,
\ee
where $Q$ can be seen as an \textit{effective} cosmological constant, except that it is not constant, but rather a spacetime function. Note that, despite not being constant, $Q$ nonetheless generates dark energy with an equation of state $w=-1$. Clearly, the Bianchi identities ensure that $\kappa \nabla^a T_{ab} + \nabla_a Q = 0$, so any variation in $Q$ must be compensated for in $T_{ab}$. This provides a natural framework for an energy/momentum exchange between dark energy and dark matter. While models incorporating interactions with baryonic matter are possible, they would be tightly constrained by observations~\cite{Amendola:1999er}.

In this work, we constrain one such model where $Q$ is linearly proportional to the energy density of CDM~\cite{deCesare:2021wmk}:
\begin{align}\label{Eq:TransferModel}
    Q  &=-\Lambda_{\rm f}+\kappa \epsilon \rho_c~,\\
    \delta Q&=\kappa \epsilon\, \overline{\rho}_c\delta_c~.\label{eq:deltaQ}
\end{align}
Here, $\Lambda_{\rm f}$ is an integration constant, $\epsilon$ is a dimensionless constant measuring the strength of the transfer from dark matter to dark energy, $\rho_c$ is the energy density of cold dark matter, and $\delta_i = \delta \rho_i / \bar \rho_i$. This is one of the simplest models of energy transfer between the dark sectors, and it is, in fact, equivalent to generalized dark matter (GDM) models~\cite{Hu:1998kj,Kopp:2016mhm,Thomas:2016iav,Ilic:2020onu} with a constant equation of state~\cite{deCesare:2021wmk}. We assume that only cold dark matter interacts with dark energy, so taking the divergence gives
\be 
\kappa \nabla^b T_{ab}^{\rm \scriptscriptstyle CDM} + \nabla_a Q = 0~.
\ee
Note that energy is transferred from dark matter to dark energy for the case $\epsilon > 0$, and in the reverse direction for $\epsilon < 0$~. There is no transfer at all for $\epsilon=0$, in which case $Q = {\rm const}$~.

This energy transfer modifies some of the equations of motion, both for the cosmological background,
\begin{align}
{\cal H}^2 &=\frac{a^2}{3}\lf(\kappa \overline{\rho} - \epsilon \kappa \overline{\rho}_c + \Lambda_{\rm f} \rg)~,\label{Eq:BackgroundMDE1}\\
{\cal H}^2+2 {\cal H}^{\prime}&=-a^2 \lf(\kappa\bar{p} + \epsilon \kappa \overline{\rho}_c - \Lambda_{\rm f} \rg)~,\label{Eq:BackgroundMDE2}\\
(1-\epsilon) \overline{\rho}_c^{\prime}+3{\cal H}\overline{\rho}_c&= 0~,
\end{align}
and for the perturbations, where, in terms of Fourier modes in the synchronous gauge,
\begin{align}
    \delta_c'=&-\frac{1}{1-\epsilon}\left(\theta_c+\frac{h'}{2}\right)~,\label{eq:perturb1}\\
    \theta_c'=&-\frac{1-4\epsilon}{1-\epsilon}\mathcal{H}\theta_c+\epsilon k^2 \delta_c~.\label{eq:purturb2}
\end{align}

In addition to dark matter (treated as pressureless dust), $T_{ab}$ receives contributions from baryons and radiation (photons and neutrinos), while, as explained above, $Q$ is the source of dark energy. The remaining equations of motion are given in Appendix~\ref{Sec:AppendixA}. The dynamics of tensor perturbations are unaffected, and the only deviations from $\Lambda$CDM appear in the scalar sector.

Many interacting dark energy models with $w\neq-1$ are affected by a large-scale instability, which gives rise to a runaway growth of the velocity perturbation of dark energy, $\theta_{\rm\scriptscriptstyle DE}$~\cite{Valiviita:2008iv}. However, since in this framework dark energy has $w=-1$, it is not necessary to define its four-velocity or its perturbation $\theta_{\rm\scriptscriptstyle DE}$~, since the four-velocity does not appear in the equations of motion (the contribution to the right-hand side of \eqref{Eq:EffectiveEFE} from dark energy is independent of its four-velocity). In such a case, the instability results obtained in Ref.~\cite{Valiviita:2008iv} no longer necessarily apply, and, in particular, the model considered here is free of this instability~\cite{deCesare:2021wmk}, and we expect this will continue to be true for a large class of functions $Q$.

Note that the family of IDE models we consider here assumes a particular scalar function $Q$, which is identified with $-\Lambda_{\rm eff}$~\cite{deCesare:2021wmk}. 
This convention differs from the standard literature, where $\tilde{Q}$ (adding a tilde to avoid confusion) instead denotes the interaction between CDM and dark energy; the relation is $\tilde{Q} = \overline{Q}^{\prime}/\kappa$. For common choices in the literature, such as $\tilde{Q} = H$, $Q$ would be an integral of $H$ over time and be non-local in time.

Furthermore, in the cases studied so far, the requirement that there should be no gradient instabilities on sub-horizon scales often determines the sign of the coupling. For this model, this requirement gives $\epsilon>0$~\cite{deCesare:2021wmk}, implying that $\overline{Q}^{\prime}<0$: energy must flow from dark matter to dark energy.

Lastly, we note that the trace-free Einstein equations \eqref{Eq:TraceFreeEFEs} may also be derived from an action principle by imposing the additional requirement of invariance under volume-preserving diffeomorphisms~\cite{Percacci:2017fsy}. This is known as unimodular gravity, but it comes with an additional condition, the so-called unimodularity constraint, that fixes the volume element to a non-dynamical background quantity. In this study, following~\cite{deCesare:2021wmk}, we work with the field equations \eqref{Eq:TraceFreeEFEs} without imposing the unimodularity condition, thus retaining full gauge invariance. For technical aspects concerning the relation between unimodular gravity and trace-free Einstein gravity, specifically concerning gauge invariance, see Refs.~\cite{Gielen:2018pvk, Bengochea:2023dep} and references therein. Alternative action functionals that give the trace-free field equations \eqref{Eq:TraceFreeEFEs} have been formulated in Refs.~\cite{Henneaux:1989zc, Kuchar:1991xd}; however, due to the diffusive nature of interactions in the general framework considered here, an action principle formulation for IDE models based on the trace-free Einstein equations may not be available.

\section{Methodology and data}\label{Sec:Methodology}

We analyzed the posterior distributions obtained using the publicly available Bayesian analysis code \texttt{COBAYA}, which was developed based on its predecessor \texttt{CosmoMC}, while a fast-dragging algorithm was implemented for efficient oversampling of the fast parameters—usually experimental nuisance parameters~\cite{Torrado:2020dgo,2019ascl.soft10019T}. \texttt{COBAYA} allows us to combine the modified cosmological theory and experimental likelihoods with the Markov Chain Monte Carlo (MCMC) sampling algorithm. 
To calculate the predictions of the model under consideration, we exploit a modified version of the Cosmic Linear Anisotropy Solving System code (\texttt{CLASS})~\cite{Blas:2011rf}. In our modified code, we updated the dark matter perturbation equations (according to \eqref{eq:perturb1} and \eqref{eq:purturb2}) and the gravitational field equations (according to \eqref{eq:field1} and \eqref{eq:field2}) to be in concordance with the IDE model in this work. The dynamical equations for the cosmological background and perturbations in the synchronous gauge are reviewed in Appendix~\ref{Sec:AppendixA}. 
The parameters being sampled are the $\epsilon$ parameter, denoting the interaction, and the six parameters from the baseline $\Lambda$CDM model, namely, the baryon energy density $\Omega_\mathrm{b} h^2$, the cold dark matter energy density $\Omega_\mathrm{c} h^2$, the angular size of the sound horizon at recombination $\theta_\mathrm{s}$, the optical depth at reionization $\tau$, the primordial scalar power spectrum amplitude $A_s$, and the spectral index $n_\mathrm{s}$. Since the $\epsilon<0$ regime is not of interest in this work due to its gradient instability, we assume a flat prior on $\epsilon \in [0,0.1]$. In Table~\ref{table:priors}, we list the priors applied to the parameters in our model.

The datasets used in this work are as follows:
\begin{itemize}
    \item The full \textit{Planck} 2018 temperature and polarization likelihoods~\cite{Planck:2018nkj,Planck:2019nip,Planck:2018vyg}, together with the \textit{Planck} 2018 lensing likelihood~\cite{Planck:2018lbu}. We refer to this dataset as \textbf{Planck2018}.
    
    \item The full \textit{Planck} 2018 temperature and polarization likelihoods~\cite{Planck:2018nkj,Planck:2019nip,Planck:2018vyg}, together with the \textit{Planck} 2018 lensing likelihood~\cite{Planck:2018lbu}, combined with DESI Baryon Acoustic Oscillation (BAO) distance measurements from galaxies and quasars~\cite{DESI:2024uvr,DESI:2024lzq,DESI:2024mwx}. For each sample at its effective redshift $z$, either both the transverse comoving distance $D_M/r_d$ and Hubble horizon $D_H/r_d$ or the angle-averaged distance $D_V/r_d$ is measured by DESI, where distance measurements are scaled by $r_d$. The full list of samples is given in Ref.~\cite{DESI:2024mwx}. We refer to this dataset as \textbf{Planck2018+DESI}.
    
    \item The full \textit{Planck} 2018 temperature and polarization likelihoods~\cite{Planck:2018nkj,Planck:2019nip,Planck:2018vyg}, together with the \textit{Planck} 2018 lensing likelihood~\cite{Planck:2018lbu}, combined with the Pantheon+ likelihood~\cite{Brout:2022vxf}. The Pantheon+ likelihood is obtained by analyzing 1701 light curves from 1550 Type Ia supernovae. We refer to this dataset as \textbf{Planck2018+SNIa}.
    
    \item The full \textit{Planck} 2018 temperature and polarization likelihoods~\cite{Planck:2018nkj,Planck:2019nip,Planck:2018vyg}, together with the \textit{Planck} 2018 lensing likelihood~\cite{Planck:2018lbu}, combined with DESI Baryon Acoustic Oscillation (BAO) data~\cite{DESI:2024uvr,DESI:2024lzq,DESI:2024mwx} and the Pantheon+ likelihood~\cite{Brout:2022vxf}. We refer to this dataset as \textbf{Planck2018+DESI+SNIa}.
\end{itemize}

A modified version of the Gelman-Rubin statistic, $R - 1$, is adopted in \texttt{COBAYA} to measure chain sampling convergence~\cite{Torrado:2020dgo,Lewis:2013hha}. We set the criterion for chain convergence at $R - 1 < 0.05$.

\begin{table}[htb]
    \centering
    \begin{tabular}{c|c}
       \hline\hline
       \textbf{Parameter}  & \textbf{Prior} \\
       \hline
        $\Omega_\mathrm{b} h^2$ & $[0.005, 0.1]$\\
        $\Omega_\mathrm{c} h^2$ & $[0.001, 0.990]$\\
        $\tau_\mathrm{reio}$ & $[0.01, 0.80]$\\
        $n_\mathrm{s}$ & $[0.8, 1.2]$\\
        $\log(10^{10} A_\mathrm{s}) $ & $[1.61, 3.91]$ \\
        $100\theta_\mathrm{s}$ & $[0.5, 10]$ \\
        $\epsilon$ & $[0, 0.1]$\\
        \hline
    \end{tabular}
    \caption{Flat priors on the parameters sampled in the MCMC analysis.}\label{table:priors}
\end{table}

\section{Results}\label{Sec:Results}

\begin{figure}[htb]
    \centering
    \includegraphics[width=\textwidth]{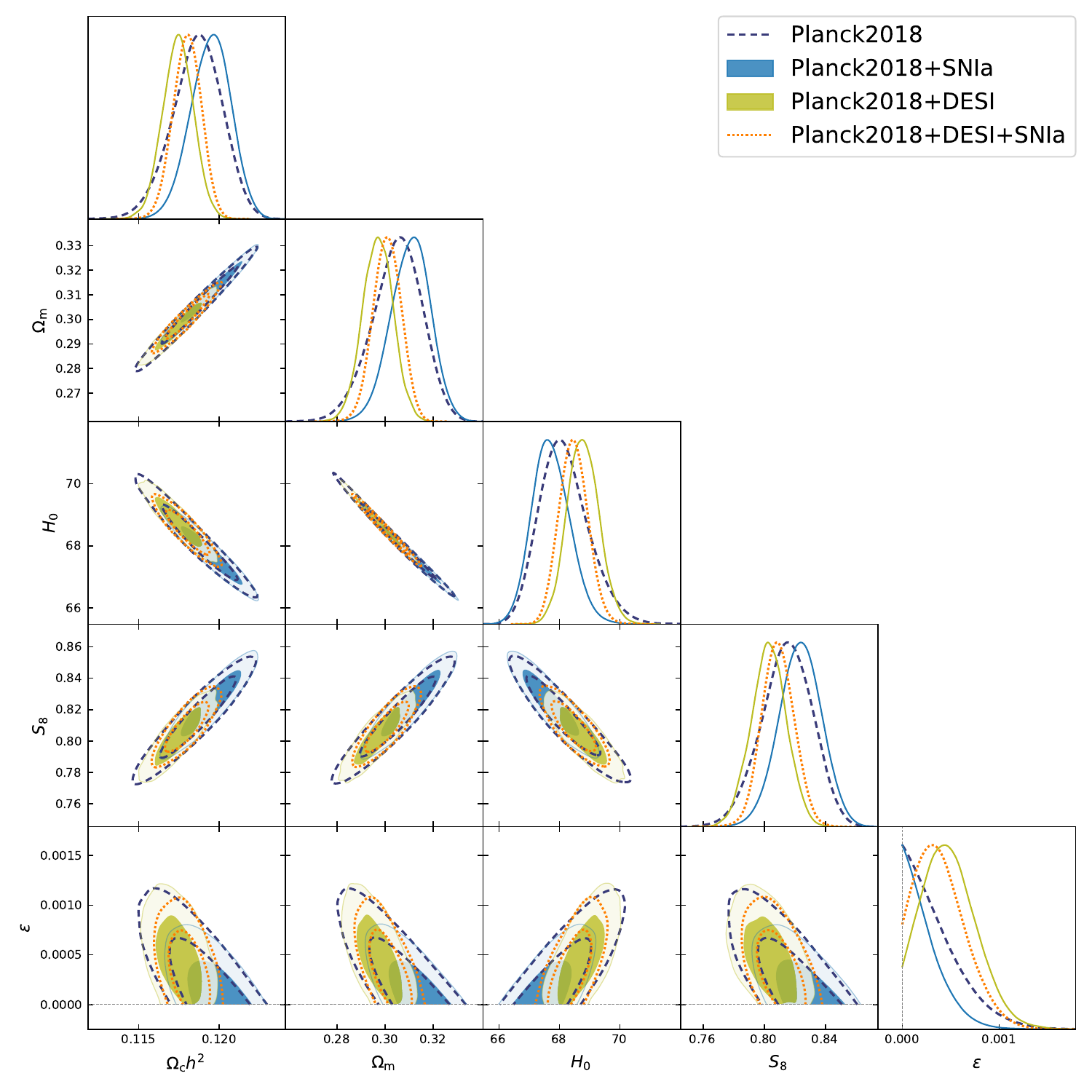}
    \caption{By imposing a uniform prior $[0,0.1]$ on $\epsilon$, the MCMC analysis of the interacting dark energy model is shown above.}\label{fig:triangle}
\end{figure}

\begin{figure}[htb]
    \centering
    \includegraphics[width=0.45\linewidth]{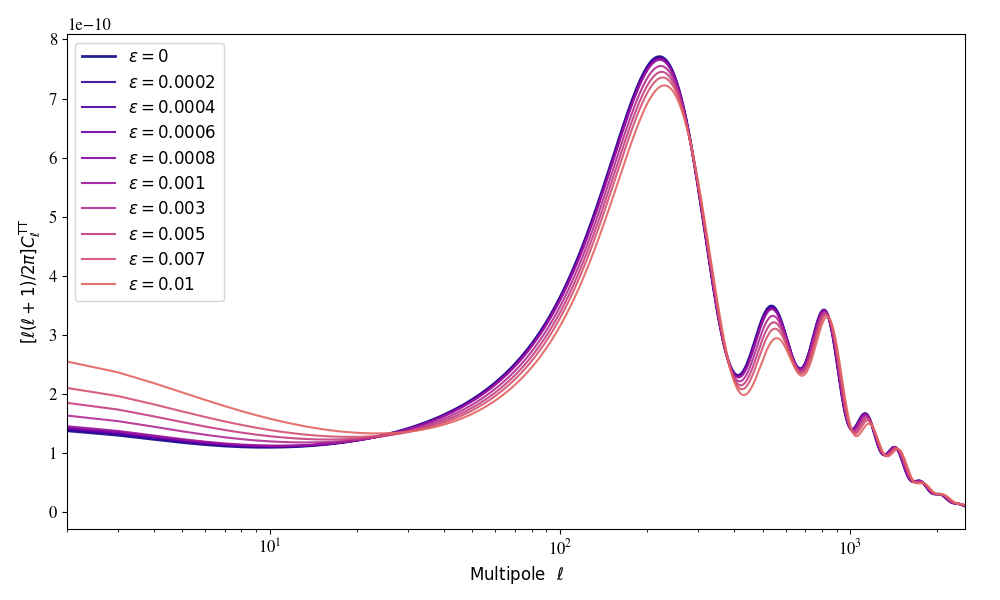}
    \includegraphics[width=0.45\linewidth]{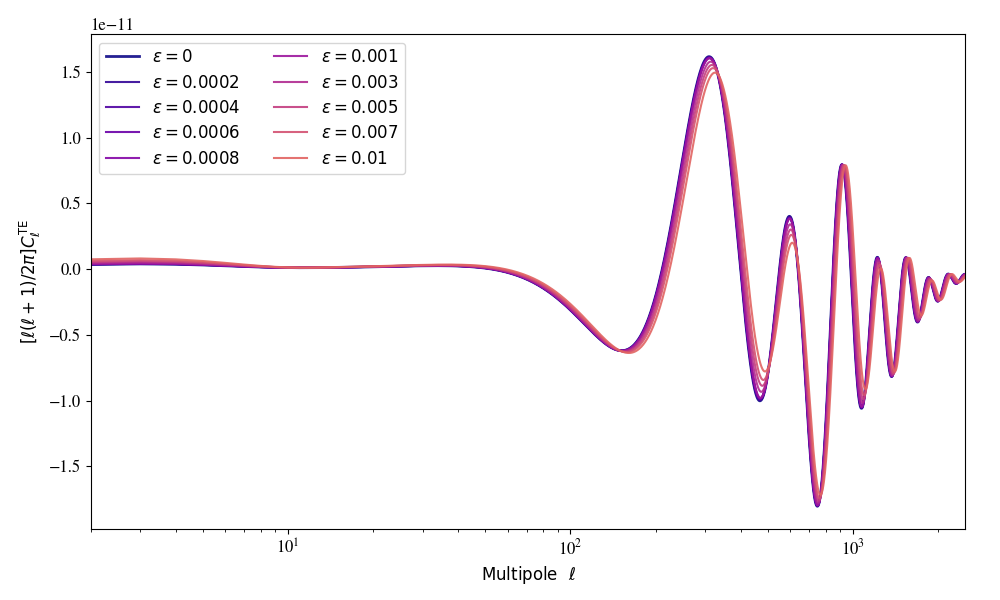}
    \includegraphics[width=0.45\linewidth]{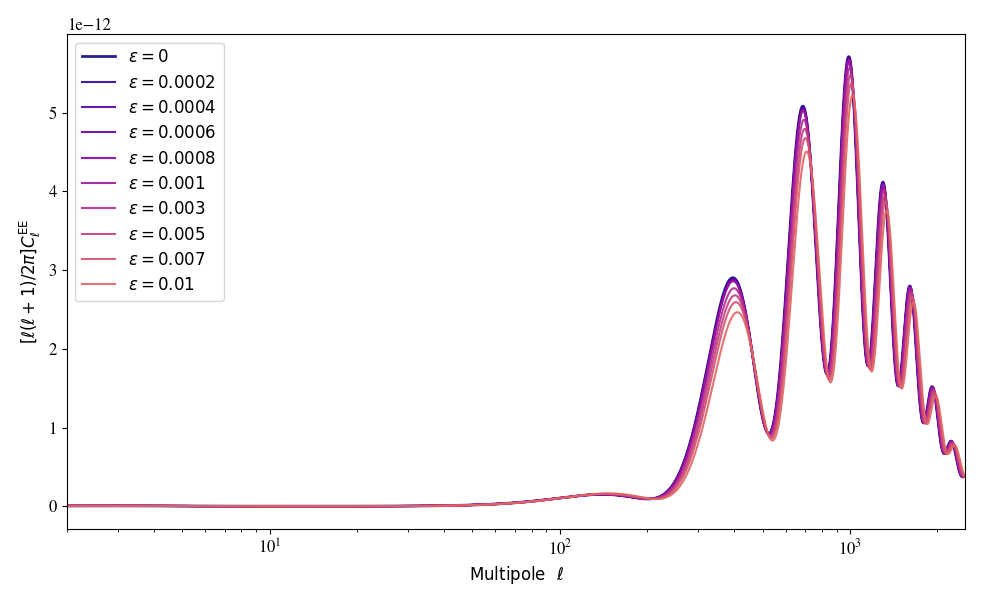}
    \caption{The theoretical predictions on the CMB TT, TE, EE power spectra for various $\epsilon$ values. }
    \label{fig:clTT}
\end{figure}

The constraints at the $68\%$ ($95\%$) confidence level on the parameters of the present model are presented in Table~\ref{table}. 
Comparing our results with those obtained by assuming a $\Lambda$CDM model for Planck2018 only, we observe a shift of about $1\sigma$ in the cold dark matter density, and consequently in the total matter density, towards smaller values. This occurs because $\epsilon$ is negatively correlated with these parameters, as we can see in Fig.~\ref{fig:triangle}. As a result, due to the very accurate measurements of $\Omega_mh^2$ from the CMB peaks, $\epsilon$ is positively correlated with $H_0$, which also shifts towards higher values by approximately $1\sigma$.
The key feature of the results is that the scale of the dimensionless interaction parameter $\epsilon$ is small, on the order of $\sim \mathcal{O}(10^{-4})$, and consistent with no interaction for Planck2018 only, with $\epsilon<0.0009$ at 95\% CL. This is expected because the interaction in this model is proportional to the energy density of dark matter, $\rho_c$, as shown in Eq.~\eqref{eq:deltaQ}. 
In the literature, interacting dark energy models with similar $\rho_c$-dependence exhibit comparably tight constraints on their interaction parameters. Their posterior distributions are concentrated close to zero, as analyzed using various combinations of datasets~\cite{Kumar:2016zpg}. The main reason for such small derived values is that, as shown in Fig.~\ref{fig:clTT}, introducing a stronger interaction in the cosmological model reduces the amplitudes of the peaks. These peaks are very well constrained by data, as this region has very small error bars. Meanwhile, the predicted Sachs-Wolfe plateau at low multipoles increases; however, this lies in the cosmic variance-limited region, where the error bars are large and the effect is less significant. 
The decrease in peak amplitudes is consistent across the CMB TT, TE, and EE spectra. The \textit{Planck} mission measures these peaks with high precision, leaving very little tolerance for such mechanisms. This effect is very similar to the constraints on the Hubble constant, where a larger $H_0$ value also leads to smaller peaks. In Fig.~\ref{fig:triangle}, the correlation between $\epsilon$ and $H_0$ can be examined more straightforwardly.

\begin{figure}[htb]
    \centering
    \includegraphics[width=0.5\linewidth]{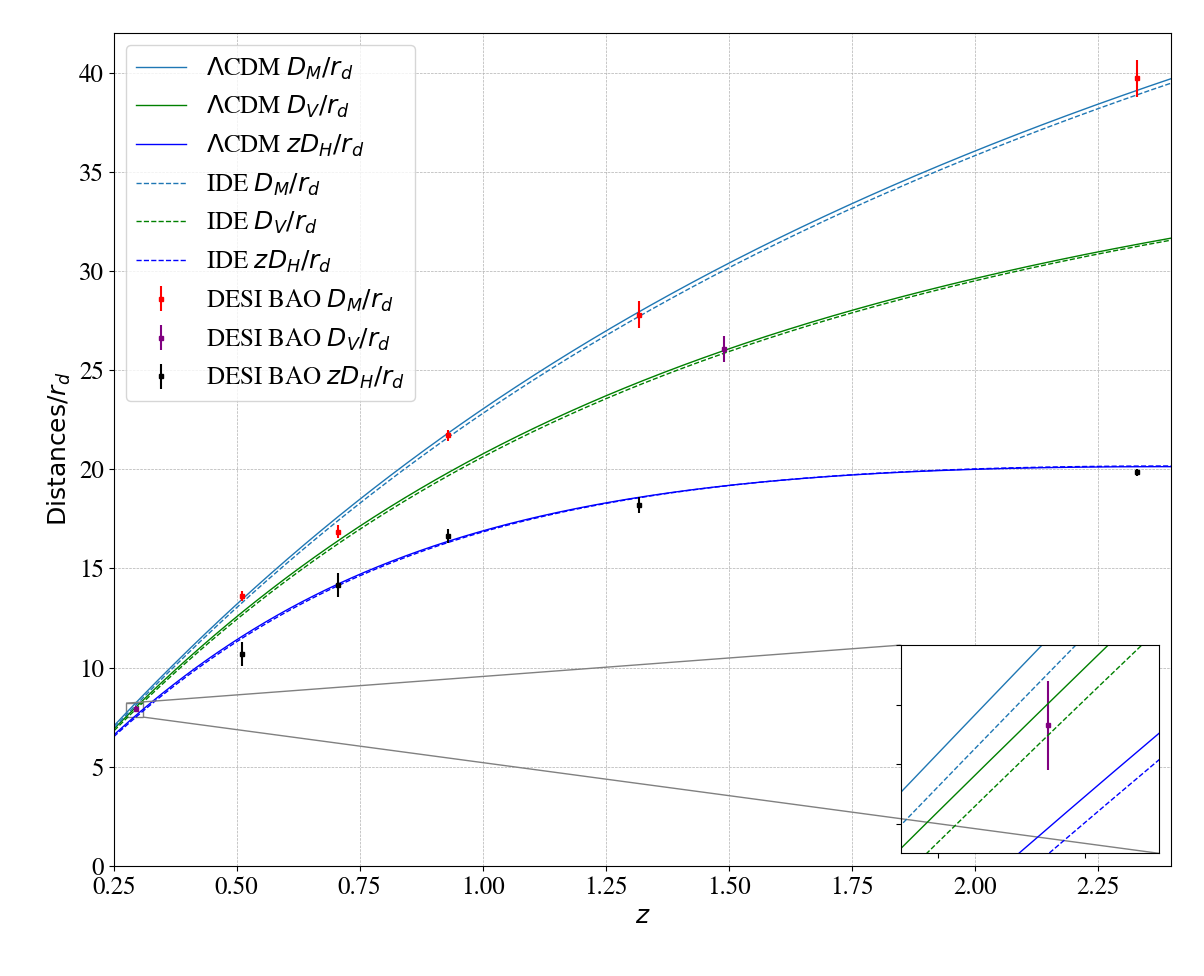}
    \caption{Distance-redshift relations from the \textit{Planck}2018+DESI dataset, adopting the IDE ($\Lambda$CDM) best-fit, are shown with dashed (solid) lines. The y-axis is rescaled by the sound horizon at baryon drag, $r_d$. For visual convenience, $D_H/r_d$ is further rescaled by $z$. The sample points represent DESI 2024 BAO distance measurements, with corresponding $\pm 1\sigma$ uncertainties. From low to high redshifts, these plotted samples include BGS at $0.1<z<0.4$, LRG at $0.4<z<0.6$, LRG at $0.6<z<0.8$, a combined sample of LRG and ELG at the overlapping redshift range $0.8<z<1.1$, ELG at $1.1<z<1.6$, quasars at $0.8<z<2.1$, and quasars at $1.77<z<4.16$.}
    \label{fig:distances}
\end{figure}

A different effect is observed in the DESI-combined analyses, which prefer a higher $\epsilon$ value. This is due to the preference of DESI 2024 BAO measurements for lower values of $\Omega_m$, which breaks degeneracies in the parameter space, slightly favoring higher values of the interaction parameter: $\epsilon=0.00049^{+0.00022}_{-0.00033}$ at 68\% CL for Planck2018+DESI and $\epsilon=0.00041^{+0.00015}_{-0.00035}$ at 68\% CL for Planck2018+DESI+SNIa. 
Compared to CMB results, DESI data alone favor a smaller total matter density when assuming a constant dark energy equation of state. In the $\Lambda$CDM scenario, the derived total matter densities from Planck2018+DESI and DESI alone are $\Omega_m = 0.3069\pm0.0050$ and $0.295\pm0.015$, respectively. This preference for a smaller $\Omega_m$ in DESI is mainly driven by the DESI distance measurements at low redshift bins ($z \leq 1.4$), more explicitly by the bright galaxy sample (BGS), luminous red galaxies (LRG), and emission line galaxies (ELG), as shown in Fig.~\ref{fig:distances}~\cite{DESI:2024mwx}. 
It is interesting to note that the LRG and ELG tracers at the same redshift bins are cross-correlated in their clustering analyses, in which BAO measurements from other redshift bins are not involved. Moreover, the DESI-combined datasets show a strong preference for a time-varying dark energy equation of state at levels $\geq 2.5 \sigma$~\cite{DESI:2024mwx,DESI:2024hhd}, while in this work based on the trace-free Einstein equations $w=-1$.
Overall, the DESI 2024 BAO data break degeneracies present in CMB analyses, such as the $\Omega_m$-$H_0$-$w$ degeneracy, leading to a decreasing trend in $\Omega_m$. When an interaction between dark sectors is allowed, this degeneracy-breaking effect induces significant changes in the posterior distribution, resulting in a nonzero $\epsilon$.

For exactly the opposite reason, i.e., because they prefer a higher value for the matter density, SNIa data are more in agreement with no interaction at all. The upper limit on $\epsilon<0.0006$ at 95\% CL for Planck2018+SNIa is stronger than in the Planck2018 case alone, and for the same reason, the Hubble constant shifts towards a lower value, making it more consistent with the $\Lambda$CDM case. 
However, as explained above, the compromise found with DESI BAO data indicates an interaction different from zero at $1\sigma$, even when SNIa data are included in the full dataset combination.

Due to correlations between $\epsilon$, $H_0$, $\Omega_m$, and the assumptions on the dark energy equation of state, it is also interesting to compare with interacting dark energy models in the literature with energy transfer from DM to DE and their effects on the tensions~\cite{Sabogal:2024yha,Benisty:2024lmj,Pan:2019jqh}. 
In the $\Lambda$CDM scenario, CMB data suggest a relatively large $\Omega_m$. When CMB data are involved, the energy-momentum transfer from dark matter to dark energy due to their interaction can naturally reduce $\Omega_m$, leading to a larger current expansion rate and, consequently, mitigating the $H_0$ tension. For the same reason, if late-time measurements such as DESI 2024 BAO are combined—where a smaller $\Omega_m$ is observed—such an interaction is more supported by data analysis. Conversely, when SNIa data are included, which prefer a higher $\Omega_m$, the interaction is in disagreement with the data. 
By relaxing the assumptions on the dark energy equation of state, either by assuming a constant $w\neq -1$ or allowing $w$ to vary with time, the low-redshift tension on $\Omega_m$ is alleviated~\cite{Shajib:2025tpd}. In this case, the constraints on the interaction are loosened, and the intensity of such an interaction becomes less significant~\cite{Giare:2024ytc,Yang:2020uga,Yang:2021hxg}. 
It remains challenging to reconcile all measurements simultaneously, as the constraints on dark-sector interactions from matter clustering and the expansion rate measured by early- and late-time projects are engaged in an arm-wrestling match.

The simple model analyzed here does not address the $H_0$ and $S_8$ tensions. 
We note that, despite some similarities between the model at hand and the `anomalous diffusion' model considered in Ref.~\cite{Perez:2020cwa}, which was proposed to alleviate the Hubble tension, the latter features a step function in the transfer potential, such that the (effective) equation of state of cold dark matter undergoes a sharp transition at recombination, an effect that is absent in the present model~\eqref{Eq:TransferModel}. Extensions of the present model, including such sudden transitions, will be investigated in future work.

\begin{figure}
    \centering
    \includegraphics[width=0.4\linewidth]{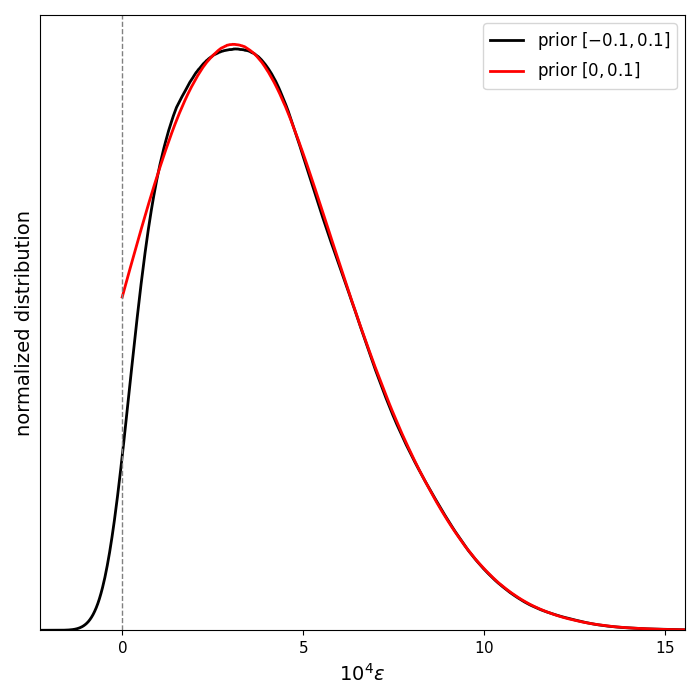}
    \caption{By adopting flat priors of [-0.1, 0.1] and [0, 0.1], the normalized $\epsilon$ posterior distributions are plotted in black and red, respectively.  The likelihoods are the Planck2018+DESI+SNIa dataset.}
    \label{fig:epsilon_posterior}
\end{figure}

Lastly, we remark that in this work, we assumed a positive flat prior for $\epsilon$ due to the gradient instability in the $\epsilon<0$ case~\cite{deCesare:2021wmk}. 
On the other hand, if we allow for negative $\epsilon$ values and assume a flat prior in the interval $[-0.1, 0.1]$, the posterior distribution features a sharp tail at the lower boundary; see Fig.~\ref{fig:epsilon_posterior}. This is consistent with the findings in Ref.~\cite{deCesare:2021wmk}, which indicate that the $\epsilon<0$ case is problematic, and it further shows that this regime is also disfavored by data.

%%%%%%%%%%%%%%%%%%%%%%%%%

\begin{table}[htb]
\renewcommand{\arraystretch}{1.5}
\resizebox{ \textwidth}{!}{
\begin{tabular}{l | c |c |c |c }
\hline\hline
Parameter & Planck2018 & Planck2018+DESI & Planck2018+SNIa & Planck2018+DESI+SNIa\\
\hline
$\Omega_\mathrm{b} h^2$  & $0.02232\pm 0.00015$ & $0.02237\pm 0.00015$& $ 0.02230\pm 0.00015$ & $0.02234\pm 0.00015$\\
$\Omega_\mathrm{c} h^2$  & $0.1186^{+0.0017}_{-0.0014}$ & $0.11744\pm 0.00095$ & $0.1195^{+0.0013}_{-0.0012}$& $0.11800\pm 0.00088$\\
$\tau_\mathrm{reio}$  & $0.0542\pm 0.0076$ & $0.0567^{+0.0069}_{-0.0077}$ & $0.0529\pm 0.0071$& $0.0560\pm 0.0071$\\
$n_\mathrm{s}$  & $0.9660\pm 0.0044$ & $0.9684\pm 0.0036$ & $0.9646\pm 0.0040$& $0.9674\pm 0.0036$\\
$\log(10^{10} A_\mathrm{s})$ & $3.043\pm 0.015$ & $3.047^{+0.014}_{-0.016}$ & $3.042\pm 0.014$& $3.046\pm 0.014$\\
$H_0$  & $68.16^{+0.67}_{-0.93}$ & $68.78\pm 0.52$ &$67.72^{+0.56}_{-0.69}$& $68.47\pm 0.47$\\
$\Omega_\mathrm{m}$ & $0.305^{+0.012}_{-0.009}$ & $0.2970\pm 0.0063$ & $0.3107^{+0.0088}_{-0.0076}$& $0.3008\pm 0.0058$\\
$\sigma_8$ & $0.8082\pm 0.0060$ & $0.8071^{+0.0058}_{-0.0064}$ & $0.8095\pm 0.0058$& $0.8079\pm 0.0058$\\
$S_8$ & $0.815^{+0.018}_{-0.016}$ & $0.803\pm 0.011$ & $0.824\pm 0.014$& $0.809\pm 0.011$\\
$\epsilon$ & $< 0.000432(< 0.000907)$ & $0.00049^{+0.00022}_{-0.00033}(< 0.000970)$ &$< 0.000293(< 0.000635)$& $0.00041^{+0.00015}_{-0.00035}(< 0.000880)$\\
\hline
\end{tabular} }
\caption{Constraints at $68\%$ ($95\%$) CL on parameters from various combinations of datasets. }\label{table}
\label{tab:tablabel}
\end{table}

\section{Conclusions}\label{Sec:Conclusions}

We have conducted a comprehensive analysis of a dark sector interaction model based on the trace-free Einstein equations where the energy-momentum transfer potential is proportional to the cold dark matter density. Using a combination of \textit{Planck} 2018 CMB, DESI 2024 BAO, and Pantheon+ SNIa data, we derived tight constraints on the interaction strength, characterized by the dimensionless parameter $\epsilon$. Our results show that $\epsilon$ is of the order of $\sim \mathcal{O}(10^{-4})$, with no significant evidence of interaction from CMB and SNIa data alone. However, when DESI data are included, a mild $1\sigma$ preference emerges for an energy-momentum transfer from dark matter to dark energy, with $\epsilon \approx 0.0004 - 0.0005$. This preference is primarily driven by the lower $\Omega_m$ values favored by low-redshift DESI BAO measurements.

Despite this indication, the interaction remains small and does not significantly alleviate the Hubble constant ($H_0$) or matter clustering ($S_8$) tensions. Nevertheless, our findings reinforce the idea that late-time interactions in the dark sector could have subtle yet measurable effects on the expansion history and structure formation. The observed correlation between $\epsilon$ and $\Omega_m$ suggests that further investigations into IDE models may yield new insights into potential modifications to the standard $\Lambda$CDM paradigm.

Trace-free Einstein gravity provides a minimal modification to general relativity that is covariant, and naturally allows a large class of IDE models. While our analysis here focused on a simple interaction model with a linear dependence on the cold dark matter energy density, a broader class of IDE models remains largely unexplored. One potential avenue for future research is the inclusion of more general transfer functions $Q$, allowing for time-dependent or scale-dependent interactions. Such extensions could potentially provide a better fit to data while simultaneously preserving key features of the standard cosmological model.

Additionally, the preference for interaction in the DESI BAO data — though not statistically strong — suggests that future large-scale structure surveys may be crucial in testing IDE scenarios. The upcoming Euclid mission, the Vera C. Rubin Observatory’s LSST survey, and future DESI data releases will significantly refine BAO and large-scale structure constraints, allowing for more precise measurements of $\Omega_m$, $H_0$, and dark sector interactions. Continued theoretical advancements and observational progress will be key to determining whether such interactions play a fundamental role in shaping the evolution of the universe.

\section*{Acknowledgements}
MdC acknowledges support from INFN iniziativa specifica GeoSymQFT. CvdB is supported by the Lancaster–Sheffield Consortium for Fundamental Physics under STFC grant: ST/X000621/1~. EDV acknowledges support from the Royal Society through a Royal Society Dorothy Hodgkin Research Fellowship. The work of EWE is supported in part by the Natural Sciences and Engineering Research Council of Canada. This work contributes to COST Action CA21136 -- Addressing observational tensions in cosmology with systematics and fundamental physics (CosmoVerse). We acknowledge IT Services at The University of Sheffield for the provision of services for High Performance Computing.

\appendix

\section{System of equations of motion}\label{Sec:AppendixA}

\renewcommand{\theequation}{A.\arabic{equation}}

In this Appendix, we review the cosmological dynamics for the class of models at hand with a general transfer potential $Q$. We specialize the results of Ref.~\cite{deCesare:2021wmk} to the synchronous gauge and re-express the equations using the notation of Ref.~\cite{Ma:1995ey}, which is convenient for implementation in the \texttt{CLASS} code. The dynamical equations for the particular model considered in this paper can be obtained by substituting Eq.~\eqref{Eq:TransferModel} into the equations shown below.

The Friedmann equations and the continuity equations, with a general transfer potential, can be split, as usual, into background and perturbative degrees of freedom. Assuming a spatially flat FLRW background, the equations of motion for the background degrees of freedom are, in conformal time,
\begin{alignat}{2}
&{\cal H}^2=\frac{a^2}{3}\lf(\kappa \overline{\rho} - \overline{Q}\rg)~,%\label{Eq:BackgroundMDE1}
\quad
&&{\cal H}^2+2 {\cal H}^{\prime}=-a^2 \lf(\kappa\bar{p} + \overline{Q}\rg)~,%\label{Eq:BackgroundMDE2}
\\
&\overline{\rho}_\gamma^{\prime}+4{\cal H}\overline{\rho}_\gamma=0~,\quad
&&\overline{\rho}_\nu^{\prime}+4{\cal H}\overline{\rho}_\nu=0~,\\
&\overline{\rho}_b^{\prime}+3{\cal H}\overline{\rho}_b=0~, \quad
&&\overline{\rho}_c^{\prime}+3{\cal H}\overline{\rho}_c=\frac{\overline{Q}^{\prime}}{\kappa}~,
\end{alignat}
where $\mathcal{H}$ is the conformal Hubble rate and primes denote derivatives with respect to conformal time. The indices are: $\gamma$ for radiation, $\nu$ for neutrinos, $b$ for baryonic matter, and $c$ for cold dark matter, while the total energy density is
\be
\overline{\rho}=\overline{\rho}_{\gamma}+\overline{\rho}_{\nu}+\overline{\rho}_{b}+\overline{\rho}_{c}~,
\ee
and the total pressure is
\be
\overline{p}=\frac{1}{3}\lf(\overline{\rho}_{\gamma}+\overline{\rho}_{\nu}\rg)~.
\ee

For scalar perturbations, we work in the synchronous gauge and use the same conventions as in Ref.~\cite{Valiviita:2008iv}, where the line element is 
\be
\de s^2 = a^2 \Big( - \de\tau^2+[(1-2\psi)\delta_{ij} +2E_{,ij}]\de x^i \de x^j  \Big)~.
\ee
The general case has been worked out in Ref.~\cite{deCesare:2021wmk}, from which the equations shown below can be obtained by imposing the gauge conditions $B=\phi=0$. In Fourier space, we have the following correspondence with the variables $h$, $\eta$ used in Ref.~\cite{Ma:1995ey}:
\be
\psi=\eta~,\quad E=-\frac{1}{2k^2}\left( h+6\eta\right)~.
\ee
In terms of these variables, the gravitational field equations read
\begin{align}
2k^2\eta-{\cal H}h'&=-a^2\left(\kappa \delta\rho-\delta Q 
\right)~\label{eq:field1},\\
k^2\eta'&=\frac{\kappa}{2}a^2(\overline{\rho}+\overline{p}) \theta~\label{eq:field2},\\
h''+2{\cal H}h'-2k^2\eta&=-3a^2\left(\kappa \delta p+\delta Q\right)~,\\
h''+6\eta''+2{\cal H}(h'+6\eta')-2k^2\eta&=-2\kappa a^2 k^2\pi~.
\end{align}

We assume that the matter fields considered here have no intrinsic entropy perturbations, that is, $c_{s\,A}^2=c_{a\,A}^2=w_A$. Then, the perturbed continuity equations (including Thomson scattering and neglecting the contribution of photons to the anisotropic stress) are
\begin{alignat}{2}
&\delta^{\prime}_\gamma+\frac{4}{3}\theta_\gamma+\frac{2}{3}h^{\prime}  = 0 ~,\quad
&&\delta^{\prime}_\nu+\frac{4}{3}\theta_\nu+\frac{2}{3}h^{\prime}   = 0 ~,\\
& \delta^{\prime}_b+\theta_b+\frac{h'}{2}  = 0 ~,
&& \delta^{\prime}_c+\theta_c+\frac{h'}{2}  = \frac{\delta Q^{\prime}-\bar{Q}^{\prime}\delta_c}{\kappa\bar{\rho}_c}~,\\
&\theta^{\prime}_\gamma-\frac{k^2}{4}\delta_\gamma=\tau_c^{-1}  (\theta_b - \theta_\gamma) ~,
&&\theta_\nu^{\prime}-\frac{k^2}{4}\delta_\nu+k^2\sigma_\nu =0 ~,\\
& \theta_b^{\prime}+{\cal H}\theta_b-k^2 c_{s,b}^2\delta_b=\lf(\frac{4\bar{\rho}_\gamma}{3\bar{\rho}_b}\rg)\,\tau_c^{-1} (\theta_\gamma-\theta_b)~,\qquad
&&\theta^{\prime}_c+{\cal H} \theta_c =\frac{k^2 \delta Q-\bar{Q}^{\prime}\theta_c}{\kappa\bar{\rho}_c}~.
\end{alignat}
where $\tau_c=(a n_e \sigma_{ T})^{-1}$ and $\sigma_{\nu}=\left(2 k^2/3(\bar{\rho}_\nu+\bar{p}_\nu)\right)\pi_\nu=\left(k^2/2\bar{\rho}_\nu\right)\pi_\nu$.
The equation for $\sigma_{\nu}$ is obtained from the quadrupolar moment of the Boltzmann equation~\cite{Ma:1995ey} (we neglect the neutrino octopole term, following Ref.~\cite{Valiviita:2008iv})
\be
\sigma_\nu^{\prime}=\frac{4}{15}\theta_\nu~.
\ee

\bibliography{biblio}

\end{document}